\documentclass[apj]{emulateapj}
\usepackage{graphicx}

\newcommand\ii{{\sc ii}}
\newcommand\iii{{\sc iii}}

\shorttitle{Towards A Removal of Temperature Dependencies from Abundance Determinations}
\shortauthors{Croxall et al.}

\begin{document}  

\title{Towards A Removal of Temperature Dependencies from Abundance Determinations:  NGC~628\thanks{Herschel is an ESA space observatory with science instruments provided by European-led Principal Investigator consortia and with important participation from NASA.}}
\author{Kevin V. Croxall$^{1,2}$, J.D. Smith$^2$, B.~R. Brandl$^3$, B. A. Groves$^{4}$, R. C. Kennicutt$^{5}$, K. Kreckel$^{4}$, 
B. D. Johnson$^{6}$, E. Pellegrini$^2$, K. M. Sandstrom$^{4,14}$, F. Walter$^4$,
L. Armus$^{7}$, P. Beir\~{a}o$^8$, D. Calzetti$^{9}$,  D.~A. Dale$^{10}$, M. Galametz$^{5}$, J. L. Hinz$^{11}$, L. K. Hunt$^{12}$, J. Koda$^{13}$, E. Schinnerer$^4$}
\affil{$^1$Department of Astronomy, The Ohio State University, 4051 McPherson Laboratory, 140 W 18th Ave., Columbus, OH, 43210
\\$^2$Department of Physics \& Astronomy, University of Toledo, 2801 W Bancroft St, Toledo, OH 43606
\\$^3$Leiden Observatory, Leiden University, P.O. Box 9513, 2300 RA Leiden, The Netherlands
\\$^4$Max-Planck-Institut f¬ur Astronomie, K\"{o}nigstuhl 17, D-69117 Heidelberg, Germany
\\$^5$Institute of Astronomy, University of Cambridge, Madingley Road, Cambridge CB3 0HA, UK
\\$^6$Institut dÕAstrophysique de Paris, UMR7095 CNRS, Universit\'{e} Pierre \& Marie Curie, 98 bis Boulevard Arago, 75014 Paris, France
\\$^7$Spitzer Science Center, California Institute of Technology, MC 314-6, Pasadena, CA 91125, USA
\\$^8$Observatoire de Paris, 61 avenue de lÕObservatoire, 75014 Paris - France
\\$^9$Department of Astronomy, University of Massachusetts, Amherst, MA 01003, USA
\\$^{10}$Department of Physics \& Astronomy, University of Wyoming, Laramie, WY 82071, USA
\\$^{11}$MMT Observatory, University of Arizona, 933 N Cherry Ave, Tucson AZ 85721
\\$^{12}$INAF - Osservatorio Astrofisico di Arcetri, Largo E. Fermi 5, 50125 Firenze, Italy
\\$^{13}$Department of Physics and Astronomy, SUNY Stony Brook, Stony Brook, NY 11794-3800, USA
\\$^{14}$Marie Curie Fellow
}
\email{croxall@astronomy.ohio-state.edu}

\begin{abstract}
The metal content of a galaxy is a key property for distinguishing between viable galaxy evolutionary scenarios, and it strongly influences many of the physical processes in the interstellar medium.  An absolute and robust determination of extragalactic metallicities is essential in constraining models of chemical enrichment and chemical evolution, however, current gas phase abundance determinations from optical fine-structure lines are uncertain to ~0.8 dex as conversion of these optical line fluxes to abundances is strongly dependent on the electron temperature of the ionized gas.  In contrast, the far-IR emission lines can be used to derive an O$^{++}$ abundance that is relatively insensitive to temperature, while the ratio of the optical to far-IR lines provides a consistent temperature to be used in the derivation of an O$^+$ abundance.  We present observations of the [O III] 88 \micron\ fine-structure line in NGC~628 that were obtained as part of the KINGFISH program.  These data are combined with optical IFU data to derive oxygen abundances for seven H\ii\ regions.  We find the abundance of these regions to all lie between the high and low values of strong line calibrations and in agreement with estimates that assume temperature fluctuations are present in the H\ii\ regions. 

\end{abstract} 
 
 
 \section{Introduction} 
Numerous aspects of the universe are governed by the abundance of elements heavier than He.   Metals provide effective means to alter the opacities, and can dominate heating and cooling, and thus determine the fundamental structure of both the interstellar medium (ISM) and stars.  Thus, extragalactic abundances can be used to distinguish between viable galaxy evolutionary scenarios. 

Despite this strong dependence on abundance, the calibration of nebular abundances from nebular emission lines remains uncertain at the level of $\sim$0.8 dex \citep{kewel}.  Elemental abundances of extragalactic H\ii\ regions are primarily determined from strong collisionally-excited optical emission lines.  This requires knowledge about the electron density ($n_{\rm e}$), the electron temperature ($T_{\rm e}$), and the fraction of atoms in each ionic state.  Detection of the temperature sensitive [O {\sc iii}] $\lambda$4363 line, in addition to the strong $\lambda$5007 line, enables a \emph{direct} \citep{dinerstein1990} determination of the electron temperature \citep{agn3}.  Unfortunately, the [O {\sc iii}] $\lambda$4363 line is inherently weak, making its detection difficult.  Detection of this line is further complicated by the fact that its strength decreases as metallicity increases, as the presence of metals causes the temperature of an H\ii\ region to decrease.  This makes [O {\sc iii}] $\lambda$4363 virtually undetectable in cooler, metal rich H\ii\ regions that are typical in spiral galaxies \citep[e.g., see][]{stasinska2005}. As a result, well-constrained direct calibrations of the abundance scale are limited to metal poor H\ii\ regions. 

In the absence of a direct measure of the temperature from the oxygen lines, abundances must be determined from \emph{strong line} techniques \citep[e.g.][]{dopita2000} which rely on either empirical calibrations \citep[i.e., ][]{PT05} or photoionization models \citep[i.e., ][]{KK04}.  These calibrations bring with them their own uncertainties and assumptions. These uncertainties can arise from atomic data (e.g., ionic cross sections, collision strengths, recombination coefficients) and the physical description of the model (e.g., gas distribution, chemical composition, dimensionality).  For example, \citet{stasszcz} show that the specific dust model adopted in a photo-ionization code can alter the temperature structure of the nebula; this results in differences in the inferred temperature between 30--60\%.  Indeed, even the manner in which observations are compared to the models can affect how line ratios are interpreted.  For example, \citet{Morisset2009} demonstrate the importance of aperture matching in reproducing the observations of nebular emission lines.  These uncertainties result in abundance scales that can differ up to 0.8 dex for a given strong line observation \citep{kewel}.  While the absolute scale of these strong-line calibrations may disagree, they can provide more reliable relative abundances.  However, each calibration suffers from its own bias effects, making it difficult to compare published results from different studies or to establish an absolute scale.

Even in the case where the temperature sensitive line is detected, there are systematic issues related to the temperature structure of the HII region.  H\ii\ regions are not simple spherical clouds uniformly filled with gas at a single temperature.  \citet{peimbert} demonstrated that fluctuations in temperature within an H\ii\ region can lead to an incorrect temperature being deduced from observations of temperature sensitive lines.  This would suggest that abundances determined via the \emph{direct} method underestimate the abundance of the ionized gas \citep{peimbert}.  Abundances in extragalactic H\ii\ regions rely on observations of strong collisionally excited lines.  In nearby bright H\ii\ regions, optical recombination lines are also detected.  These different sets of lines, optical recombination and collisionally excited, yield significantly different temperatures for the same cloud \citep[e.g.,][]{esteban09,garciarojas}.  Indeed, different elements also indicate different temperatures for a given ionization zone.  A more controversial solution was proposed by \citet{nicholls}, who recently suggested that these disagreements may be resolved by relaxing the assumption that electrons are in a Maxwellian energy distribution.  

In addition to the optical forbidden lines, two strong [O \iii] fine-structure transitions exist in the far-infrared (far-IR), at 52$\mu$m and 88$\mu$m.  As the metal content of an H\ii\ region increases, and the temperature decreases, contributions to cooling from far-IR fine-structure lines become more important \citep{stasinska2002}.  Around an oxygen abundance of half of the solar value
\footnote{We have assumed the solar oxygen abundance of \citet{asplund}, 12+log(O/H) = 8.69$\pm$0.05.} 
(12$+$log(O/H) $\sim$ 8.4),  [O \iii] cooling becomes dominated by emission in the far-IR.  Due to low excitation temperatures, if O$^{++}$ exists in a nebula, then the far-IR lines will be emitted.  Moreover, as these lines are much less sensitive to the temperature of the nebula, they will be largely insensitive to temperature fluctuations whose existence, within and across an H\ii\ region, may lead to the inaccurate abundance determinations \citep[e.g.,][]{peimbert}.  Thus observation of these far-IR lines may enable a calibration of the metallicity scale that is largely independent of the temperature uncertainties that plague current calibration schemes.

While inaccessible from the ground, both [O \iii] 88 and 52 \micron\ lines were previously observed by the Infrared Space Observatory (ISO) \citep{kessler}.  \citet{brauher} show that in many nearby galaxies, the emission from the [O \iii] 88\micron\ line is between 0.03\%\ -- 2\%\ of the total far-IR dust continuum.    \citet{garnetiso} used ISO observations of the 88~$\mu$m and narrow band imaging of [O~\iii] 5007\AA\ to derive T[O \iii] in M51.  However, they were unable to derive an absolute oxygen abundance as 1) they lacked a direct measure of H$^{+}$, 2) they did not observe [O~\ii], and 3) multiple H\ii\ regions were contained within the large, $\sim$80\arcsec, ISO beam.  Studies of the galactic ISM were able to focus on single H\ii\ regions which had measurements of H$^+$ from the radio continuum.  Still the absence of infrared emission lines of O$^+$ resulted in a reliance on ionization correction factors \citep{simpsoniso}.  Given that the fractional abundance of O$^{++}$ was less than 0.5, these abundances are quite uncertain.  Other authors simply reported their derived abundance of O$^{++}$ as a lower limit on the total oxygen abundance \citep{herniso}.  The determination of oxygen abundances from ISO is further complicated because we must deduce the relative number of oxygen to hydrogen atoms within the observed aperture.  Given the large beam, this becomes difficult, as contributions from diffuse H\ii\ can surely contaminate large sections of imaging data that would be required to deduce the amount of H$^+$ in the ISO beam associated with the H\ii\ regions.

In short, ISO's spatial resolution meant that 1) extragalactic H\ii\ regions could not be isolated, 2) abundance gradients within even nearby galaxies could not be studied, and 3) matching far-IR observations to optical data was not possible.  Consequently, no other ionization states of oxygen could be folded in to yield a total oxygen abundance from ISO data.  

With the Photodetector Array Camera \& Spectrometer (PACS) onboard the {\it Herschel Space Observatory} we are for the first time able to target the far-IR [O \iii] lines, among other neutral and ionized fine structure lines, on sub-kiloparsec scales in nearby galaxies.  The {\sc Kingfish} project \citep[Key Insights on Nearby Galaxies: a Far- Infrared Survey with Herschel, ][]{kingfish} is an open-time Herschel key program investigating critical dust and gas diagnostics within nearby galaxies that span a wide range of key physical parameters \citep{kingfish}.  We are able to further constrain the physical properties of the H\ii\ regions by combining the far-IR observations of the {\sc Kingfish} program with matched mid-IR observations available from the {\it Spitzer} Infrared Nearby Galaxies Survey \citep[{\sc Sings},][]{sings}.   Development of optical Integrated Field Units (IFU) enables us to also obtain optical spectral maps.  With these maps, we can extract spectra from matched apertures rather than attempting to scale long slit spectra to the far-IR observations.  The spectral coverage and spatial resolution of these combined observations allow us to use the temperature insensitive far-IR [O \iii] fine-structure lines to investigate abundance patterns in nearby galaxies.

In this paper we present a pilot investigation of H\ii\ regions in NGC~628 by combining data from optical IFU spectroscopy with spectroscopic observations from the {\sc Kingfish} and {\sc Sings} programs.  NGC~628 was selected for this as investigation due to the existence of publicly available PPAK data, which was obtained as part of the  PPAK Integral Field Spectroscopy (IFS) Nearby Galaxies Survey \citep[PINGS,][]{pings}.  In \S2 we present the observations and discuss the data processing.  The various temperature and density dependancies are discussed and abundances are derived in \S3.  Finally, we summarize the results of this study in \S4.

\section {Observations}
\subsection{Far-IR PACS Maps}
\begin{figure}[tp] 
\epsscale{1.2}
   \centering
   \plotone{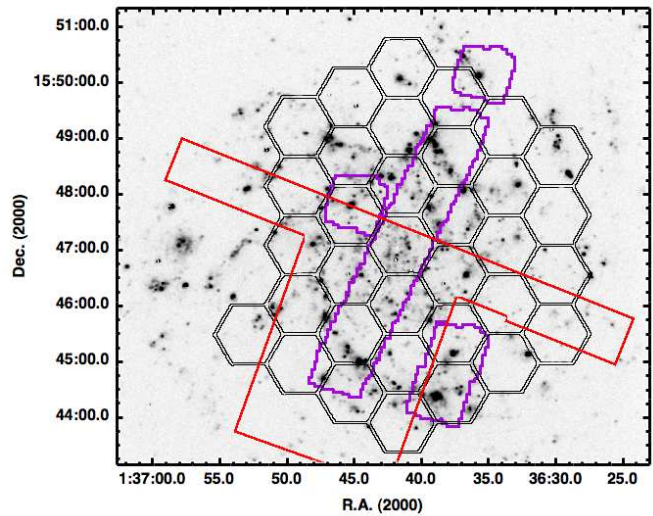}
   \caption{Footprints of PINGS, KINGFISH, and IRS observations of NGC~628 are shown in black, purple, and red, respectively, overlaid on the H$\alpha$ image from SINGS.  The hexagonal pattern seen in the PINGS footprint is the shape to the PMAS IFU, each tile represents an individual pointing.  IRS observations are a combination of a long strip from SINGS and a wider strip covering the south.  The PACS data from KINGFISH were assembled from pointings of the $\sim47$\arcsec\ field-of-view that were observed as a strip and individual extranuclear regions.  Offsets between PINGS observations of NGC~628 were corrected by aligning the peak H$\alpha$ emission in individual hexagonal tiles of the PINGS map with the SINGS H$\alpha$ map.}
   \label{fig:aors}
\end{figure}

Using PACS onboard Herschel we carried out far-IR spectral observations of the [O~\iii] 88$\mu$m line within NGC~628 as part of the {\sc Kingfish} Open Time Key Program \citep{kingfish}.  We report a detailed summary of observations in Table \ref{t:obs}.  The IFU onboard PACS has a roughly 47\arcsec$\times$47\arcsec\ field of view composed of 25 9\farcs4 $\times$ 9\farcs4 spatial elements \citep{pacs}.  In Figure \ref{fig:aors} we show footprints of the spectral observations of NGC~628.  Far-IR observations were designed to be aligned with existing observations with the {\it Spitzer} Infrared Spectrograph \citep[IRS,][]{houck}.   

All PACS spectral observations were obtained in the Un-Chopped Mapping mode and reduced using the Herschel Interactive Processing Environment (HIPE) version 8.0 \citep{hipe}.  Reductions applied the standard spectral response functions, flat field corrections, and flagged instrument artifacts and bad pixels  \citep[see][]{pacs,kingfish}.  The dark current, determined from each individual observation, was subtracted during processing as it was not removed via chopping.  In-flight flux calibrations were applied to the data.  These calibrations resulted in absolute flux uncertainties on the order of 15\% with relative flux uncertainties between each {\it Herschel} pointing on the order of $\sim$10\%.  

\begin{figure}[bhp] 
\epsscale{1.2}
   \centering
   \plotone{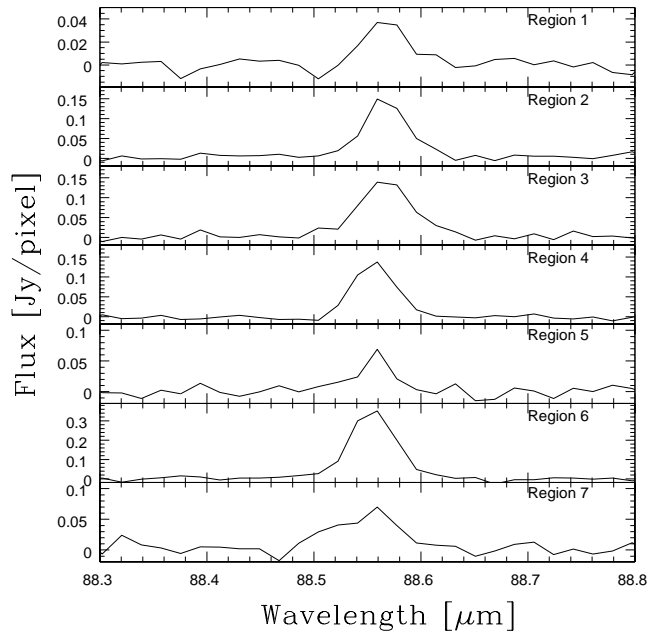}
   \caption{Extracted PACS spectra of the [O~\iii] 88~\micron\ line for each of the 7 regions presented in this study.}
   \label{fig:pacs}
\end{figure}

Initial processing assumed that {\it Herschel's} baseline was stable for the duration of the observation.  However baseline drifts and long term instrumental transients do occur causing the non-astrophysical continuum, which is dominated by emission from {\it Herschel} itself, level to drift.  These transients were strongly correlated with motions of the grating and of {\it Herschel}.   
Indeed, modeling of the spectral energy distributions of KINGFISH galaxies \citet{dale} has indicated that less than 2\% of the spectral continuum  was astrophysical.
Thus, we have assumed that continuum emission from dust is undetected, and used observations of the continuum region next to the [O~\iii]~88~\micron\ line to correct for transients.  This has significantly improved our ability to detect line emission.  The averages of the clean off-observations obtained were subtracted from observations to correct for the thermal background contributed by {\it Herschel}.  Subsequently, all spectra within a given spatial element were combined.  Final spectral cubes with 2.85\arcsec\ spatial pixels were created by projecting dithered raster positions of each spectral line together.  The final [O~\iii] map is shown overlaid on the SINGS H$\alpha$ image in Figure \ref{fig:linemapsb}.  Extracted PACS spectra are shown in Figure \ref{fig:pacs} for each of the regions labeled in Figure \ref{fig:linemapsb}.  

\subsection{PPAK Data}
NGC~628 was observed with the Potsdam Multi-Aperture Spectrophotometer (PMAS) \citep{roth} using the PMAS fiber Package (PPAK) IFU as part of the PINGS Survey \citep{pings}.  By tiling observations of NGC~628, PINGS has produced emission line maps for several strong optical lines \citep{pings628}, see Figure \ref{fig:aors}.  Initial registration of emission line maps of NGC~628 was performed by matching stellar sources to optical images taken as part of the SINGS survey \citep{pings628}.  However, this registration did not account for offsets that existed between the different pointings (Rosales-Ortega, private communication).  To account for these offsets, we have re-registered the PINGS emission line maps of NGC~628, on a tile-by-tile basis, with H$\alpha$ imaging from SINGS \citep{singsIMAGING}.  

\begin{figure*}[htp] 
   \centering
   \epsscale{1.2}
   \plotone{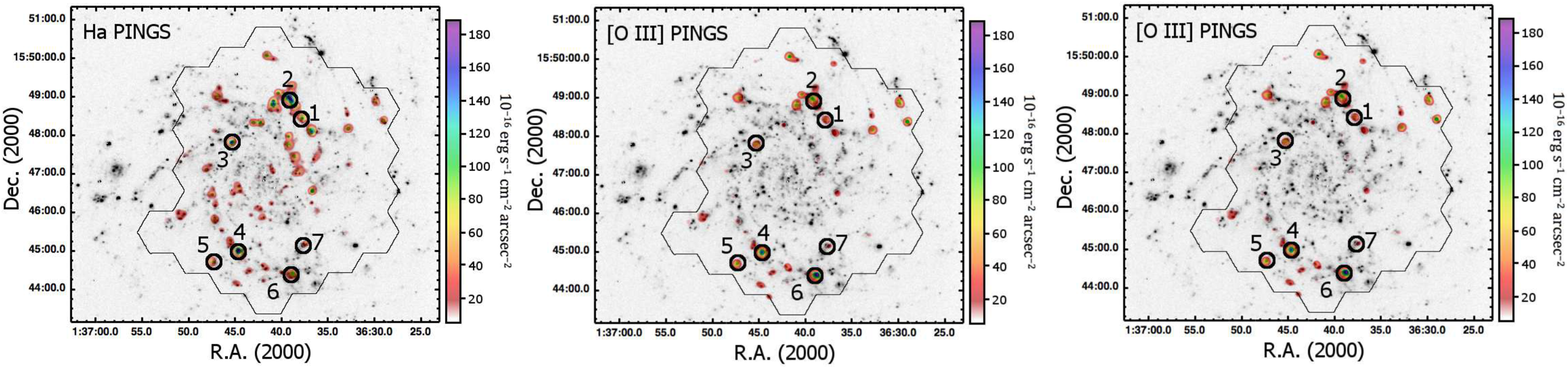}
     \caption{Spectral maps of PINGS H$\alpha$ (left) and [O~\iii]~5007~\AA\ (center) convolved to match the {\it Hercshel} beam at 88 \micron\  and KINGFISH [O~\iii] 88 \micron\ emission (right) in NGC~628 all overlaid on the continuum subtracted SINGS H$\alpha$ image.  We number the regions which were extracted for abundance determinations.  We required that regions were robustly detected in both PINGS and KINGFISH spectral datasets for consideration in this study.  Note that the extraction size was chosen to be larger than the beam size, shown in the lower corner of the [O~\iii] plot, to ensure there was not extended emission associated with each H\ii\ region which was missed by the extraction.  Footprints of the mapped area are denoted by the thin black lines.}
   \label{fig:linemapsb}
\end{figure*}

Given that we are working with both optical and far-IR observations, it is essential that we convolve the data to a matched resolution. The individual fibers of PPAK are 2\farcs7 in diameter, and thus require convolution to ensure a valid comparison of fluxes from the disparate data sets.  Observations of calibration stars show that the PSF produced by PPAK is well matched by a top-hat profile.  A convolution kernel was made following the work of \citet{kernel} to match a  2\farcs7 diameter top hat to the PSF of PACS at 88 \micron\ ($\sim$9\arcsec).  The convolved PINGS spectral mapping of H$\alpha$ is shown in Figure \ref{fig:linemapsb}

PINGS observations of NGC 628 were only dithered in the central pointing, leaving the remaining tiles (including those containing our regions) with incomplete spatial coverage. To correct for this incomplete spatial coverage (i.e., aperture bias), we use SINGS narrow-band H$\alpha$ images \citep{singsIMAGING}. From these narrow-band images we determined the H$\alpha$ flux of our regions, corrected for NII contamination, and used the ratio of this flux to the spectroscopic H$\alpha$ flux to scale all the line fluxes measured from the PINGS spectroscopy. These scale factors ranged between 5 -- 30 with a typical value of $\sim$10.
While other species of ionized gas may have distributions that are different from H$^+$, these differences are likely to be small relative to the size of the apertures used. The correction for [N~\ii] contamination of the H$\alpha$ images was determined using the PINGS spectroscopy, from which we derived [N~\ii]/H$\alpha$ between 0.24 and 0.50, with an average value of 0.37 for the regions investigated in this paper.  This is similar to the average value of 0.345$\pm$0.046 derived by \citet{kenn09} and the value of 0.408$\pm$0.168 derived from the driftscan observations of \citet{moustakas}.

Dust Reddening estimates were determined by comparing the observed ratios of Balmer emission line fluxes to intrinsic ratios. Ratios of intrinsic Balmer line strength were interpolated from the tables of \citet{caseb} for Case B recombination (H$\alpha/H\beta$ = 2.86) using the extinction curve of \citet{odonnell}.  We assume that the flux from the [O~\iii]~88~\micron\ line is un-extinguished.  Reddening corrected line fluxes relative to H$\beta$ are reported in Table \ref{t:lines} for the 7 HII regions analyzed in this paper as indicated in Figure \ref{fig:linemapsb}.\\
~\\~\\

\subsection{IRS Data}
NGC~628 was also observed by the {\sc Sings} program \citep{sings} with {\it Spitzer}-IRS \citep{houck}.  
The observed wavelength range includes the [Ne~\ii] and [Ne~\iii] lines at 12.8 and  15.6~\micron\, respectively, and the [S~\iii] lines at 18.7 and 33.5~\micron.  These lines can be used to constrain the metallicity, ionization state, and density of an H\ii\ region.  
Both the nuclear and extra-nuclear regions were targeted by the short-low (SL) module.  In addition to SL observations, the long-low (LL) module targeted strips across each target galaxy (see Figure \ref{fig:aors}).  Additional SL and LL data were obtained for NGC~628 by {\it Spitzer} Program JDSMITH\_PBM/20518.   For a more detailed description of the {\it Spitzer} data, including reduction techniques, we refer the reader to \citet{pahfit}.  

{\it Spitzer}-IRS spectra were extracted using CUBISM \citep{cubism}.  Broad spectral features and the mid-IR lines were fit and decomposed using PAHFIT \citep{pahfit}.  We fixed the silicate absorption optical depth at zero as \citet{pahfit} found that negligible amounts of absorption from silicate dust were needed to reproduce the equivalent widths of lines in kpc scale regions of IRS spectral observations of NGC~628.  We only accept data with a surface brightness larger than three times the root mean square (RMS) noise as physically meaningful detections.    We note that, unfortunately, the hydrogen (7-4) transition at 12.36$\mu$m was not seen in the {\it Spitzer}-IRS data, leading us to rely on optical lines to determine the H$^+$ ions.

\subsection{H\ii\ Region Sample}
We have used 20\arcsec\ apertures, which, at the distance of NGC~628, $\sim$9~Mpc \citep{dist1, dist2}, correspond to an aperture approximately 900~pc in diameter; these apertures are shown in Figure \ref{fig:linemapsb} This size of aperture will likely encompass several individual H\ii\ regions that are all located within the same region of a given spiral arm.  We have selected this large aperture due to (1) the beam size of {\it Herschel}, $\sim$8\arcsec; (2) the incomplete filling factor of the PINGS spectral map; and (3) to ensure the inclusion of extended emission associated with each H\ii\ region. Line ratios for each region are reported in Table \ref{t:lines}, while derived quantities are reported in Table \ref{t:der}.

\section{Results}
We will show that the 88~\micron/H$\beta$ line ratio alone can provide a much tighter constraint on the O$^{++}$ abundance than the optical line ratio (5007~\AA/H$\beta$), particularly in a high-metallicity environment where temperature sensitive lines are difficult to detect. With additional constraints on density, from [SII]~6716/6731~\AA\ and [SIII]~18/33~\micron\ ratios, and on temperature, from optical to far-IR line ratio, we can obtain even a tighter constraint on the abundance.

\subsection{Dependencies in Determining Abundances}

\begin{figure}[bt] 
   \centering
   \epsscale{1.2}
   \plotone{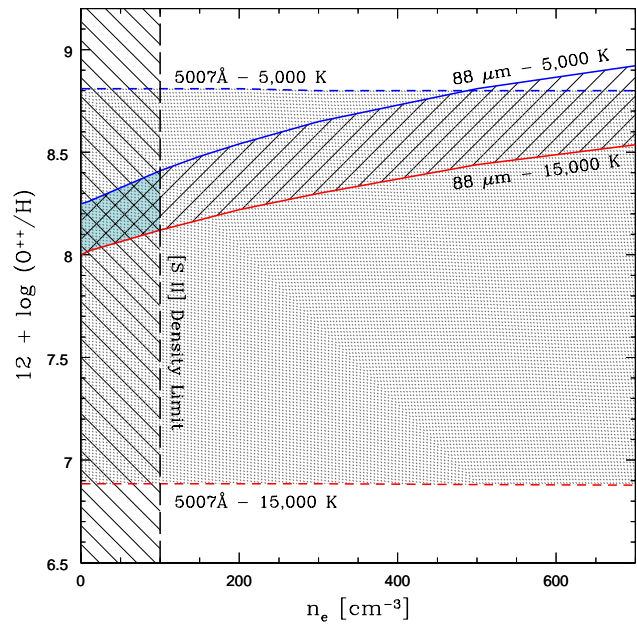}
   \caption{The plausible range of O$^{++}$ abundance (here in units of 12 + log(O$^{++}$/H)) as a function of electron density ($n_{\rm e}$) for the average values of 5007\AA/H$\beta$ and 88$\mu$m/H$\beta$ in our selected H\ii\ regions in NGC~628.  The horizontal dashed lines show the possible range of abundances that can be derived from the ratio of 5007\AA/H$\beta$, whereas the solid lines bound the possible range of abundances given the 88 $\mu$m far-IR emission line.  The [S~\ii] low density limit is indicated by the vertical long-dashed line, as all regions are in this limit, the density is constrained to lie to the left of this line.  The final range of abundances for this set of line ratios, given no information about the temperature, is highlighted in light blue, at the union of all the other regions.  Note that the possible range of 12+log(O$^{++}$/H$^{+}$) is greatly reduced with the addition of the far-IR fine-structure line.}
   \label{fig:comp}
\end{figure}

Far-IR emission from  O$^{++}$ has a low excitation temperature of 163 K \citep{drainebook}.  Therefore, if O$^{++}$ exists in a typical nebula, the 88 $\mu$m fine-structure line will be present.  In Figure \ref{fig:comp}  we show the range of O$^{++}$ abundances possible given the average emission line ratios observed in NGC~628 from the de-reddened optical and far-IR spectral maps.  We have used the five level atom, via the FIVEL program \citep{fivel}, to derive emissivities for the optical (5007 \AA), and far-IR (88 $\mu$m) [O~\iii] lines and H$\beta$ at 5,000K and 15,000K.  These temperatures were chosen as they approximate the observed range of electron temperatures in H\ii\ regions.   

As can be seen, the ratio of [O~\iii] 88 $\mu$m/H$\beta$ places a stronger restriction on the possible abundance of O$^{++}$ present  compared to the ratio of [O~\iii] 5007 \AA/H$\beta$.  However, this ratio is also somewhat sensitive to the electron density  of the H\ii\ region, unlike the ratio of 5007 \AA/H$\beta$ at low densities.  We note that we use H$\beta$ relative to 88 $\mu$m as our data do not contain mid-IR hydrogen lines, such as the Humphreys-$\alpha$ line, which might otherwise be used in a measure of relative abundance.  We constrain the density via PINGS observations of the [S~\ii] 6716 / 6731\AA\ ratio.  Observations of the [S~\ii] doublet in the optical confirm that these regions are in the low-density limit \citep[see Table \ref{t:der}, e.g. $n_{\rm e}\lesssim$ 100 cm$^{-3}$,][]{drainebook}.  In regions with {\it Spitzer}-IRS coverage, the [S~\iii] 18 / 33 $\mu$m ratio confirms the low density.  

In order to establish a lower limit on the density, we calculate the emission measure, EM, determined from the H$\beta$ surface brightness, $I_{\rm{H}\beta}$,
\begin{equation} EM = \frac{4\pi~\times~I_{\rm{H}\beta}}{h\nu_{\rm{H}\beta} \times~\alpha^{\rm{eff}}_{\rm{H}\beta}}~= \int n_{\rm{H}^+} n_{\rm e}~dl,\end{equation}
where $\alpha^{\rm{eff}}_{\rm{H}\beta}$ is the effective recombination rate for H$\beta$, and $l$ is the line-of-sight length through the H\ii\ region \citep[e.g., ][]{agn3}.  We find that the EM in our sample of H\ii\ regions lies between 4$\times10^4$ -- $2\times 10^5$.  To form a  conservative estimate we assume that each H\ii\ region complex fills each 900pc aperture and has a filling factor of 0.1 \citep{filling}.  This yields lower density limits in the range 20 -- 40 cm$^{-3}$.  These density limits, 30 $\le$ $n_{\rm e}$ $\le$ 100, in turn constrain the possible abundance determined from the ratio of [O~\iii] 5007 \AA/H$\beta$.

\begin{figure*}[t] 
   \centering
   \epsscale{1.22}
   \plotone{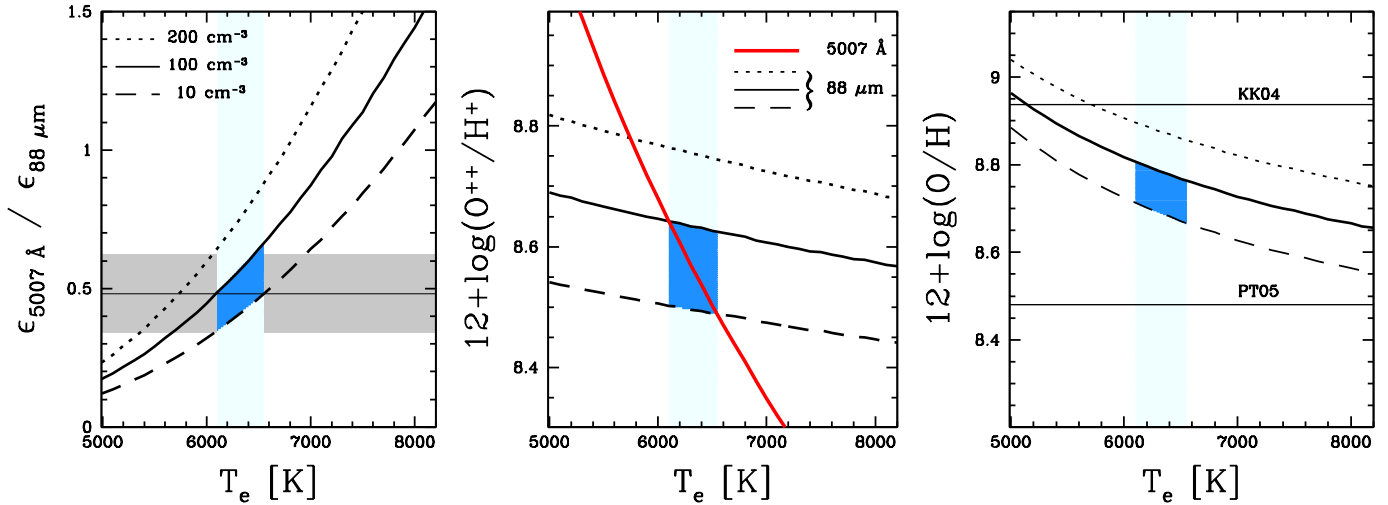}
   \caption{Temperature and abundance constraints on Region 6 (see Figure \ref{fig:linemapsb}). 
   	Left: The ratio of the 5007\AA\ and 88 $\mu$m [O~\iii] emissivities plotted as a function of electron temperature.  The different curves show the ratio at different densities; these densities are the same in all panels.  The observed flux ratio for this region is indicated by the horizontal line, while its uncertainty is shown by the grey shaded area.  
	Center: The abundance of O$^{++}$, as determined from the 88~$\mu$m (black) and 5007~\AA\ (red) lines relative to H$\beta$, is plotted as a function of temperature.	
	Right: The total oxygen abundance is shown as a function of temperature.  
	For comparison, the abundance determined from the strong optical lines using the KK04 and the PT05 calibrations are indicted by the horizontal lines.  In all panels, the vertical blue bar marks the likely temperature inferred from this line ratio.  The darker segment of the blue bar highlights the restricted parameter space given that [S~\ii] and [S~\iii] lines indicate the region is in the low density limit.}
   \label{fig:region}
\end{figure*}

\subsection{Abundances Estimates}
\subsubsection{Empirical Abundances: KINGFISH + PINGS}
Given the depth of the PINGS optical spectra, and the relatively high oxygen abundance, we are unable to accurately measure the temperature sensitive [O~\iii] 4363\AA\ line in any region of NGC~628.  We can, however, directly determine the temperature of these regions by comparing the strong lines in the optical and far-IR regimes.  We show diagnostic plots, derived from a simple 5-level atom, of this temperature determination in Figure \ref{fig:region}, left and central panels for one representative region.  While the 88 \micron\ fine-structure-line is sensitive to the electron density of the gas (see Figure \ref{fig:comp}), it can be used in conjunction with the [O~\iii] 5007\AA\ line, which has a significantly different excitation energy, to constrain the average electron temperature in the O$^{++}$ region of an H\ii\ region.  
 
 As discussed in \S2.1, we assume that the aperture correction derived for H$\alpha$, to correct the PINGS maps for incomplete sampling of the area subtended by the PINGS field-of-view, could be applied to all emission lines.  Given our large aperture and relatively extended H\ii\ regions, this is generally a reasonable assumption.  However, such a procedure can also lead to systematic errors and care must be taken.  Indeed, comparison of the peaks and profiles of the selected regions with the layout of the PMAS fibers indicates that PINGS mapping missed the central emission peak of Region 3, skewing the ratio of [O~\iii]/[O~\ii] emission.  Indeed, while this aperture exhibits a low FIR abundance given its proximity to the galaxy center, none of the strong line abundances show a similar low abundance compared to other similarly calibrated abundances.  Due to uncertainties in the WCS of PINGS fibers we are unable to assign accurate quantitative estimates of the uncertainty due to fiber placement for each region.  Indeed, it is only due to the obvious mismatch of Region 3 that this effect was detected.  While we do not see indications that other apertures were affected, it is certain that all apertures were affected by this systematic error to some extent.  
  
As previously noted, while the optical lines are much more sensitive to temperature changes compared to the far-IR [O~\iii] emission, the far-IR lines are more dependent on the density of the gas as shown in Figure \ref{fig:comp}.  Given the range of densities that fall in the low density limit, there is a systematic uncertainty of $\sim$0.05 dex in the abundance determination due to density uncertainties.
 
 \begin{figure}[tp] 
   \centering
   \epsscale{1.2}
   \plotone{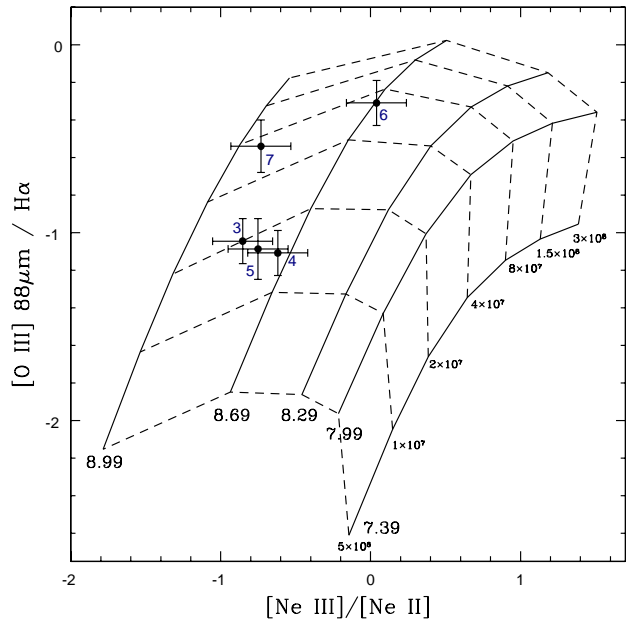}
   \caption{The ratio of [O~\iii] 88~$\mu$m / H$\alpha$ is plotted against the [Ne \ii] 12.8~$\mu$m / [Ne \iii] 15.5~$\mu$m ratio for regions with both PACS and IRS coverage (see Figure{fig:linemapsb} for identification of regions).  We constrain the ionization parameter and metallicity using by using the mid-IR Ne ratio in conjunction with [O~\iii] 88 \micron, none of which are sensitive to the temperature.  A grid of metallicity and ionization parameter from the Mappings~\iii\ photoionization code is overlaid on the data.  While the metallicity sampled by the models is sparse, we find good agreement with the empirical abundances derived using the PACS and PINGS data.}
   \label{fig:midir}
\end{figure}

One of the largest uncertainties in our FIR abundance determination is the abundance of O$^+$, as we do not have a temperature insensitive constraint on this ion.  The low temperatures indicated by the ratio of the 5007\AA\ and 88 $\mu$m [O~\iii] lines indicates that $\sim$30--60\%\ of oxygen is singly ionized, depending on the temperature assumed.  
In order to derive the O$^{++}$/O$^+$ ratio, we need to find a temperature of the O$^+$ region.  This requires us to make an assumption since we don't have a temperature constraint on this ion.  One possible assumption is the two-zone approximation \cite[e.g.,][]{pagel92} but this assumption may break down.  For example, it is not clear that empirical two-zone approximations \cite[e.g.,][]{pagel92,garnett92} are independent of temperature fluctuations; neither is it clear that these relations can be extrapolated to these cool temperatures.  \citet{oey} found that  two-zone photoionization models of H\ii\ regions in the LMC are more consistent with an isothermal nebula when T[O~\iii] $<$10,000~K.  Additionally, \citet{pil2010} found that metal-rich SDSS galaxies lie below the empirical calibrations, i.e., lower T[O~\ii] for a given T[O~\iii].  The empirical two-zone relations suggest that T[O~\ii] is greater than T[O~\iii] by approximately 1,200 K on average in the H\ii\ regions selected in this study.  Given the high metallicity and low temperatures, we have adopted the average of an isothermal nebula model where T[O~\ii] = T[O~\iii] \citep{oey} and the \citet{garnett92} temperature relation,
\begin{equation}T([O~II]) = 0.7 \times (T([O~III])^{-1} + 0.3).\end{equation}  
Adoption of either temperature model changes the derived abundances by $\sim$0.07 dex compared to the far-IR abundances reported in Table \ref{t:der}.

To summarize the sources of error we have considered in addition to random errors originating in measuring spectral lines:
\begin{itemize}
\item PPAK scaling to correct for incomplete coverage could produce a random error as large as $\sim$0.2~dex.  In most cases, the agreement with the mid-IR abundances suggest the actual error is typically lower than this, see Section 3.2.2. 
\item For our abundances, we have assumed $n_{\rm e}$ lies mid-way between the upper and lower density limits.  This could introduce an error of $\sim$0.05~dex.
\item A two-zone temperature structure following Equation 2 would lower the far-IR abundances by 0.07~dex; adoption of an isothermal temperature structure would raise the far-IR abundances by 0.07~dex
\end{itemize}

\subsubsection{Abundances from Photoionization Models: KINGFISH + SINGS}
The conversion from O$^{++}$/H$^+$ to O/H requires some constraint on the abundance of O$^+$ (or on the ratio of O$^+$/O$^{++}$).  Although we cannot directly constrain the abundance of O$^+$ independent of the temperature of the H\ii\ region we are able to use the mid-IR [Ne \iii]/[Ne \ii] ratio to constrain the local ionization parameter and metallicity.  In Figure \ref{fig:midir} we plot the ratio of [O~\iii] 88~$\mu$m / H$\alpha$ against the [Ne \iii] 15.5~$\mu$m / [Ne \ii] 12.8~$\mu$m ratio, for the five regions that have {\it Herschel} PACS, {\it Spitzer} IRS, and PPAK coverage.  We have extracted the H$\alpha$ flux from narrow band images that were corrected for [N~\ii] emission.  As the mid-IR spectral maps are fully sampled, we have used the imaging data to ensure that uncertainties due to the incomplete coverage of the PINGS maps do not influence this measurement of the metallicity.  

\begin{figure}[tp] 
   \centering
   \epsscale{1.2}
   \plotone{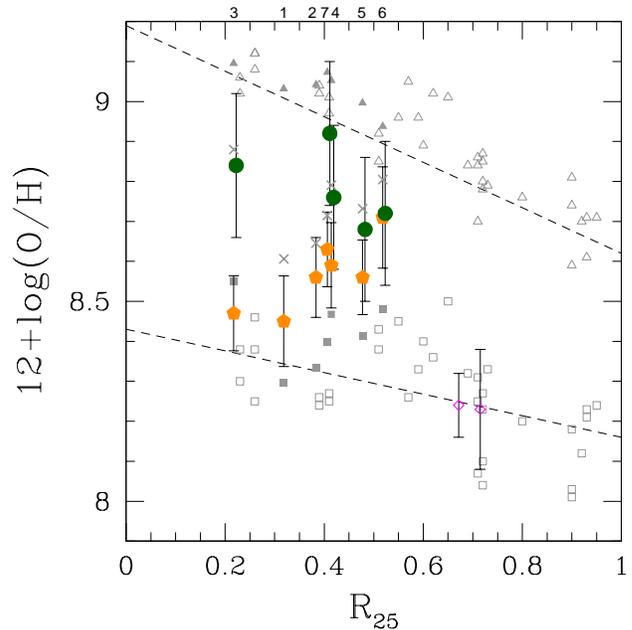}
   \caption{Radial abundance gradient in NGC~628.  Gas-phase abundances determined using the KINGFISH and PINGS data are plotted as orange pentagons while abundances determined using KINGFISH and {\it Spitzer}-IRS data are plotted as filled green circles, slightly offset in radius for plotting clarity.  Grey squares, triangles, and crosses represent PT05, KK04, PG12$^2$ 
 abundances, respectively.  Magenta diamonds indicate direct abundance determinations.  Filled points denote the seven regions selected in this study; unfilled points are taken from  observations of H\ii\ regions reported in \citet{moustakas} and \citet{cast}.  The radial gradient(s) found by \citet{moustakas} are plotted as dashed lines.  Regions are identified by their number across the top of the plot.}
   \label{fig:gradient}
\end{figure}

In Figure \ref{fig:midir}, we also overlay a photoionization model from the Mappings \iii\ photoionization code \citep{mappings3}, assuming a continuous star formation rate and an average density of 10 cm$^{-3}$.  The assumption of continuous star formation could lead to a systematic underestimate of the metallicity by $\sim$0.3 dex.  We find there is rough agreement between the metallicity indicated by the photoionization model in Figure \ref{fig:midir} and the abundance derived by scaling the [O~\iii] temperature (see Table \ref{t:der} and Figure \ref{fig:gradient}).  Indeed, the largest uncertainty in this case is the lack of a finer grid of metallicity in Mappings \iii.  We note that while Region 3 shows an anomalously low abundance (see \ref{fig:gradient}), compared to a typical gradient, when combining the KINGFISH and PINGS data, it does not appear to exhibit a discrepant abundance when using the mid-IR Ne lines.

\subsection{Comparison to Previous Estimates}
Recent mid-IR observations were used by \citet{h10103} to dispute the presence of temperature fluctuations in H10103, a metal-rich H\ii\ region in M101.  While our current data do not allow us to directly investigate the presence of temperature fluctuations in our regions,  we can compare our FIR derived abundances with the strong line calibration of \citet[][hereafter PG12]{pena}, which is based on observations of H\ii\ regions for which there is a measurement of the temperature inhomogeneity parameter, $t^2$.  

The PG12 calibration includes a correction to account for oxygen that has depleted onto dust grains.  This allows their abundances to be directly compared to stellar abundances and and models of chemical evolution which rely on total abundance and not gas phase abundances \citep{peimpeim}.  Adopting this correction increases our FIR abundances by 0.12 dex, bringing the FIR abundances into better agreement with the mid-IR abundances based on models which include dust.  We only include this correction when comparing to PG12, as this correction has not been adopted for the PT05 or KK04 calibrations.

We find that the average (excluding Region 3) FIR determined abundance is offset by 0.15, -0.46, and -0.05 dex from PT05, KK04, and PG12 respectively.  Thus, while significant offsets exist with both PT05 and KK04, we find that abundances estimated from temperature insensitive FIR lines are in agreement with the strong line abundances derived from PG12, which have accounted for an average temperature fluctuation.  

We plot the inclination corrected gas-phase radial abundance gradient of NGC~628 in Figure \ref{fig:gradient}, assuming an inclination of 6.5$^\circ$.  The FIR determined abundances, derived from simple 5-level atom physics, are plotted as orange pentagons, with PT05, KK04, and PG12\footnote{We remind the user this calibration includes a correction for depletion of oxygen onto dust} strong-line abundances for the same regions plotted as filled grey squares, triangles, and crosses respectively.  We also plot the gradient from mid-IR abundances (see Figure \ref{fig:midir}), based on more complex photoionization models, as green pentagons, for the five regions with {\it Spitzer}-IRS data.  Direct abundances reported by \citet{cast} for two H\ii\ regions in NGC~628 are also shown as magenta diamonds.  While both regions lie well below our FIR abundances, they also lie at significantly larger radius, making direct comparison with our sample  difficult. This could imply the presence of a relatively steep abundance gradient; it could also imply that temperature fluctuations are important as they would cause these direct abundance determinations to increase. Strong-line abundances from additional H\ii\ regions, compiled from the literature by \citet{moustakas}, are plotted as open grey triangles (KK04) and squares (PT05).  The random errors associated with strong-line abundances are comparatively small, $\sim0.05$ dex, as the strong-line fluxes are indeed strong.  However, such error bars do not display the inherent uncertainty of the calibration.

We do not attempt to fit an abundance gradient to NGC~628 given the restricted range of R$_{25}$ in our data sample.  As can be seen in Figure \ref{fig:gradient} the strong-line abundances for the seven FIR selected regions resides in-between the PT05 and KK04 gradients derived by \citet{moustakas} which are plotted as dashed lines.

\section {Conclusions}
Combining spatially resolved spectroscopy from {\it Herschel} PACS, {\it Spitzer} IRS, and PMAS we have derived the gas-phase oxygen abundance of several H\ii\ regions in NGC~628 using both optical and far-IR fine-structure lines.  The ratio of far-IR fine structure lines to the hydrogen recombination lines has the advantage of being nearly temperature insensitive.  This indifference of metallicity-sensitive ratios to the electron temperature is paramount to being able to unambiguously calibrate oxygen abundances from nebular emission.  While these lines have been observed previously by ISO, {\it Herschel} has enabled observations on sub-kpc scales in galaxies outside the local group.  Furthermore, advances in optical integral field unit spectroscopy make it possible to easily match the larger apertures required by observing in the far-IR.

We derive abundances for 7 regions in NGC~628 using the far-IR [O~\iii] 88 $\mu$m line in combination with optical spectroscopy.  The derived abundances lie between the KK04 and PT05 abundances reported by \citet{moustakas} for the radial position of the H\ii\ regions here studied.  Furthermore, the far-IR abundances are consistent with the PG12 strong line calibration which attempts to account for temperature fluctuations that may exist within H\ii\ regions, when accounting for depletion of oxygen onto dust grains as was considered in this calibration.  Our abundance determination utilizes the optical [O~\ii] emission to quantify the O$^+$ abundance, which retains a reliance upon the temperature determined.  Thus, we have derived an O$^{++}$ temperature consistent with the infrared observations using the ratio of [O~\iii] 88$\mu$m / 5007\AA, and converted that temperature to an appropriate O$^+$ temperature.  

Using Mappings III models \citep{mappings3}, we find a good agreement between the empirical FIR abundances and model based abundances derived using the ratio of the mid-IR [Ne~\ii] and [Ne~\iii] lines and the ratio of [O~\iii] 88$\mu$m / H$\alpha$, both of which are insensitive to temperature.  These data do not span a large range of radius.  They are consistent with slope of the gradient derived by \citet{moustakas}, assuming different intercepts, and with the absence of a gradient.
Accordingly, we do not attempt to derive an abundance gradient for NGC~628.  

NGC~628 is only one of the KINGFISH galaxies for which detections of [O~\iii] 88$\mu$m have been obtained.  Indeed, 40 out of 54 galaxies in the KINGFISH sample have detections of [O~\iii] 88 \micron, permitting similar studies spanning a wide parameter space.  We are expanding the sample of galaxies with the requisite observations that will provide matched optical spectra to selected star-forming galaxies from KINGFISH.  From a statistically significant sample of H\ii\ regions we will be able to provide a re-calibration of the strong-line abundance scales that is based on the far-IR [O~\iii].  This scale will be significantly less sensitive to the temperature of an H\ii\ region, extending  empirical calibrations to the high abundances and low temperatures that have been as yet inaccessible.

\acknowledgments
We would like to thank the referee for a quick and thoughtful report that has improved this paper.  KVC would like to thank Rick Pogge and Brett Andrews for enlightening discussions on this topic. PACS has been developed by a consortium of institutes led by MPE (Germany) and including UVIE (Austria); KU Leuven, CSL, IMEC (Belgium); CEA, LAM (France); MPIA (Germany); INAF-IFSI/OAA/OAP/OAT, LENS, SISSA (Italy); IAC (Spain). This development has been supported by the funding agencies BMVIT (Austria), ESA-PRODEX (Belgium), CEA/CNES (France), DLR (Germany), ASI/INAF (Italy), and CICYT/MCYT (Spain).  
HIPE is a joint development by the Herschel Science Ground Segment Consortium, consisting of ESA, the NASA Herschel Science Center, and the HIFI, PACS and SPIRE consortia.  
This work is based [in part] on observations made with Herschel, a European Space Agency Cornerstone Mission with significant participation by NASA. Support for this work was provided by NASA through an award issued by JPL/Caltech.  
This research has made use of the NASA/IPAC Extragalactic Database (NED) which is operated by the Jet Propulsion Laboratory, California Institute of Technology, under contract with the National Aeronautics and Space Administration.


\clearpage

\begin{deluxetable}{lcccccccccc}  
\tabletypesize{\scriptsize}
\tablecaption{PACS Spectroscopic Observations of NGC 628}
\tablewidth{0pt}
\tablehead{   
  \colhead{Obs. ID}	&
  \colhead{R.A.}	&
  \colhead{Dec.}	&
  \colhead{Raster}	&
  \colhead{Raster}	&
  \colhead{Duration}	&
  \colhead{Repe-}	&
  \colhead{t$_{\rm{[O~III]}}$}\\
  \colhead{}		&
  \colhead{(J2000)}	&
  \colhead{(J2000)}	&
  \colhead{Pattern}&
  \colhead{Step (\arcsec)}	&
  \colhead{(s)}	&
  \colhead{titions}	&
  \colhead{(s)}	}
\startdata
1342223786	&	1~36~35.50	&	+15~50~11.0	&	2x2	&	4.5 x 4.5	&	3202	&	2	&	1200	\\
1342223784	&	1~36~38.80	&	+15~44~25.0	&	2x2	&	4.5 x 4.5	&	3228	&	1	&	600	\\
1342223781	&	1~36~37.50	&	+15~45~12.0	&	2x2	&	4.5 x 4.5	&	3230	&	1	&	600	\\
1342223779	&	1~36~45.10	&	+15~47~51.0	&	2x2	&	4.5 x 4.5	&	3215	&	1	&	600	\\
1342223777	&	1~36~41.80	&	+15~47~00.0	&	13x2	&	23.5 x 4.5	&	16692	&	2	&	7800		
\enddata
\label{t:obs}
\tablecomments{Summary of PACS spectral observations of [O~\iii] NGC~628.  Units of right ascension are hours, minutes, and seconds, and units of declination are degrees, arcminutes, and arcseconds. Note that multiple lines were observed in each observation.  The duration is the full time of the observation, while t$_{\mbox{[O~\iii]}}$ is the portion of the observation devoted to the [O~\iii] line at 88$\mu$m.}
\end{deluxetable}

\begin{deluxetable}{lccccccc}  
\tabletypesize{\scriptsize}
\tablecaption{Line Fluxes Relative to H$\beta$}
\tablewidth{0pt}
\tablehead{   
  \colhead{}	&
  \colhead{Region 1}	&
  \colhead{Region 2}	&
  \colhead{Region 3}	&
  \colhead{Region 4}	&
  \colhead{Region 5}	&
  \colhead{Region 6}	&
  \colhead{Region 7}\\
  \colhead{}	&
  \colhead{+50.6, +46.9}	&
  \colhead{-55.5, +84.0}	&
  \colhead{-37.6, +112.7}	&
  \colhead{+41.7, -122.1}	&
  \colhead{-59.2, -112.4}	&
  \colhead{-40.2, -158.9}	&
  \colhead{+80.4, -138.8}
}
\startdata
$[$O~\iii$]$~88$\mu$m   & 0.85$\pm$0.07 & 1.33$\pm$0.01 & 1.36$\pm$0.03 & 1.79$\pm$0.01 & 1.63$\pm$0.21 & 2.97$\pm$0.01 & 1.62$\pm$0.13  \\
$[$O~\iii$]$~4959\AA    & 0.13$\pm$0.02 & 0.12$\pm$0.02 & 0.11$\pm$0.02 & 0.18$\pm$0.03 & 0.27$\pm$0.04 & 0.47$\pm$0.08 & 0.13$\pm$0.02  \\
$[$O~\iii$]$~5007\AA    & 0.36$\pm$0.05 & 0.39$\pm$0.06 & 0.45$\pm$0.06 & 0.54$\pm$0.08 & 0.78$\pm$0.11 & 1.44$\pm$0.20 & 0.30$\pm$0.04  \\
$[$O~\ii$]$~3727\AA     & 2.33$\pm$0.22 & 2.17$\pm$0.16 & 1.35$\pm$0.16 & 1.80$\pm$0.23 & 2.23$\pm$0.17 & 2.06$\pm$0.17 & 1.80$\pm$0.16  \\
$[$N~\ii$]$~6548\AA     & 0.28$\pm$0.03 & 0.27$\pm$0.03 & 0.37$\pm$0.04 & 0.26$\pm$0.03 & 0.20$\pm$0.03 & 0.18$\pm$0.03 & 0.34$\pm$0.03  \\
$[$N~\ii$]$~6584\AA     & 0.81$\pm$0.08 & 0.79$\pm$0.08 & 1.06$\pm$0.11 & 0.76$\pm$0.08 & 0.58$\pm$0.08 & 0.51$\pm$0.08 & 0.97$\pm$0.08  \\
$[$S~\ii$]$~6717\AA     & 0.48$\pm$0.05 & 0.48$\pm$0.05 & 0.30$\pm$0.03 & 0.41$\pm$0.05 & 0.35$\pm$0.05 & 0.31$\pm$0.05 & 0.47$\pm$0.05  \\
$[$S~\ii$]$~6731\AA     & 0.33$\pm$0.03 & 0.32$\pm$0.03 & 0.22$\pm$0.02 & 0.28$\pm$0.03 & 0.26$\pm$0.03 & 0.22$\pm$0.03 & 0.35$\pm$0.03  \vspace{.75mm}\\
H$\alpha$/H$\beta$      & 3.86$\pm$0.20 & 4.24$\pm$0.14 & 4.71$\pm$0.14 & 3.70$\pm$0.21 & 3.79$\pm$0.15 & 3.49$\pm$0.15 & 3.39$\pm$0.14
\enddata
\tablecomments{Coordinates for regions are given as offsets in seconds of arc from the center of NGC~628, 01h36m41.7s +15d47m01s}
\label{t:lines}
\end{deluxetable}

\begin{deluxetable}{lccccccc}  
\tabletypesize{\scriptsize}
\tablecaption{Derived Quantities for H\ii\ Regions}
\tablewidth{0pt}
\tablehead{   
  \colhead{}	&
  \colhead{Region 1}	&
  \colhead{Region 2}	&
  \colhead{Region 3}	&
  \colhead{Region 4}	&
  \colhead{Region 5}	&
  \colhead{Region 6}	&
  \colhead{Region 7}
}
\startdata
R$/$R$_{25}$\tablenotemark{a} & 0.32 & 0.38 & 0.22 & 0.41 & 0.48 & 0.52 & 0.41 \vspace{.5mm} \\
$[$S~\iii$]~18/33$ & -- & -- & 0.65$\pm$0.03 & 0.75$\pm$0.04 & 0.92$\pm$0.03 & 0.55$\pm$0.03 & 0.65$\pm$0.03 \vspace{.5mm}\\
$[$Ne~\iii$]/[$Ne~\ii$]~15/12$ & -- & -- & 0.14$\pm$0.01 & 0.24$\pm$0.01 & 0.18$\pm$0.01 & 1.09$\pm$0.05 & 0.19$\pm$0.01 \vspace{.5mm}\\
$T_{\rm e}$([O~\iii]) [K]\tablenotemark{b} & 5950$\pm^{450}_{600}$  & 5500$\pm^{450}_{650}$  & 5610$\pm^{450}_{450}$ & 5550$\pm^{500}_{700}$ & 6050$\pm^{500}_{600}$ & 6100$\pm^{350}_{350}$ & 5050$\pm^{700}_{5050}$ \vspace{.5mm} \\
O$^{++}$/O$^+$ & 0.50$\pm$0.03 & 0.67$\pm$0.03 & 1.24$\pm$0.03 & 1.15$\pm$0.03 & 1.10$\pm$0.03 & 2.16$\pm$0.03 & 0.76$\pm$0.03 \vspace{1.5mm}  \\  
\multicolumn{8}{c}{12+log(O/H)}  \vspace{1.mm}\\ \hline
PT05\tablenotemark{c} & 8.30$\pm$0.20 & 8.33$\pm$0.20 & 8.55$\pm$0.22 & 8.47$\pm$0.23 & 8.41$\pm$0.19 & 8.48$\pm$0.18 & 8.40$\pm$0.21 \\
KK04\tablenotemark{d} & 9.03$\pm$0.20 & 9.04$\pm$0.20 & 9.09$\pm$0.22 & 9.05$\pm$0.23 & 9.00$\pm$0.19 & 8.94$\pm$0.18 & 9.07$\pm$0.21 \\
PG12\tablenotemark{e} & 8.61$\pm$0.20 & 8.65$\pm$0.20 & 8.88$\pm$0.22 & 8.79$\pm$0.23 & 8.73$\pm$0.19 & 8.80$\pm$0.18 & 8.71$\pm$0.21 \vspace{1.mm} \\
Mid-IR & -- & -- & 8.84$\pm$0.18 & 8.76$\pm$0.18 & 8.80$\pm$0.18 & 8.69$\pm$0.18 & 8.92$\pm$0.18 \\
Far-IR\tablenotemark{f} & 8.45$\pm$0.11 & 8.56$\pm$0.10 & 8.47$\pm$0.09 & 8.59$\pm$0.09 & 8.56$\pm$0.13 & 8.71$\pm$0.09 & 8.63$\pm$0.11 
\enddata
\tablenotetext{a}{inclination corrected, R$_{25}$ = 315\arcsec\ = 13.7 kpc}
\tablenotetext{b}{Temperature assuming the [S~\ii] low density limit of 100 cm$^{-3}$}
\tablenotetext{c}{\citet{PT05}}
\tablenotetext{d}{\citet{KK04}}
\tablenotetext{e}{\citet{pena}}
\tablenotetext{f}{ An additional 0.12 dex must be added to account for depletion onto dust grains when comparing with PG12, stellar abundances, and models of chemical evolution.}
\label{t:der}
\end{deluxetable}

\section {Online Only Additional Figures}
\begin{figure*}[bp] 
   \centering
   \epsscale{1.2}
   \plotone{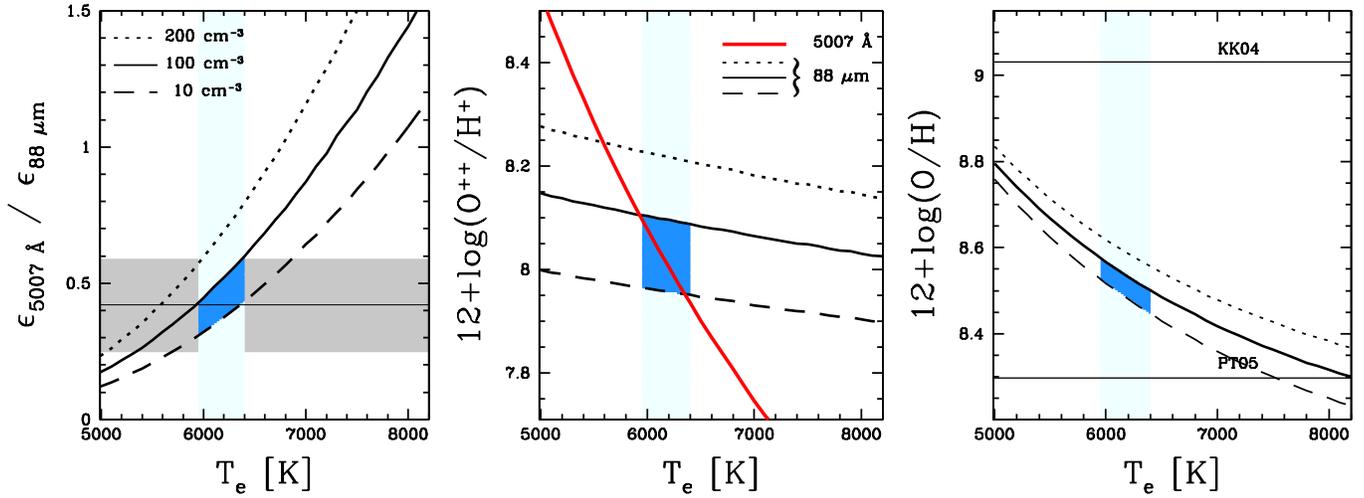}
   \caption{ Region 1.}
   \label{fig:region_online}
\end{figure*}
\begin{figure*}[bp] 
   \centering
   \epsscale{1.2}
   \plotone{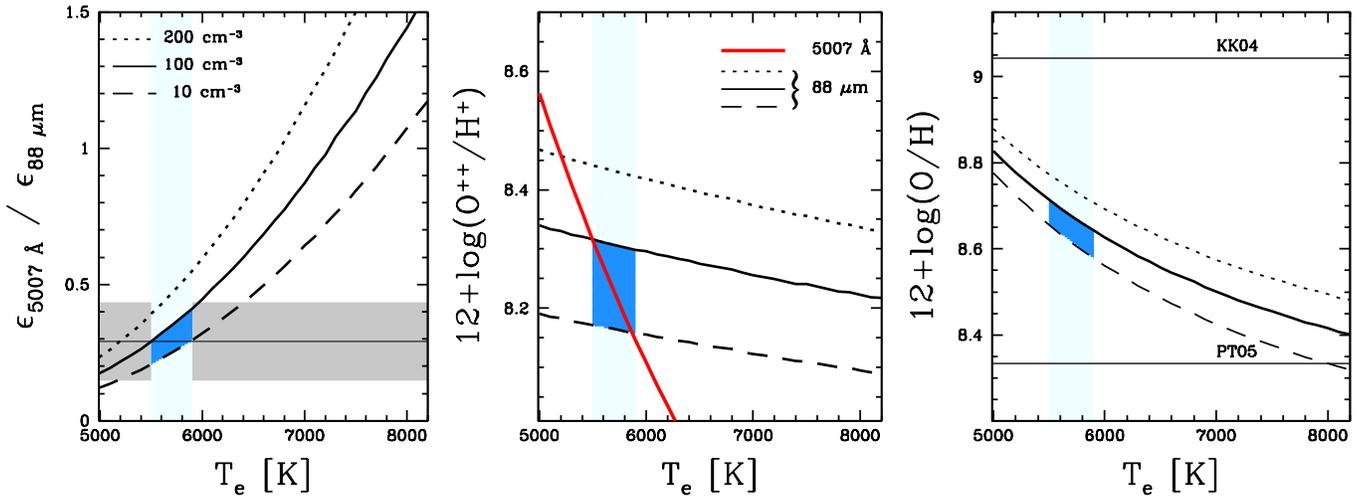}
   \caption{ Region 2.}
   \label{fig:region_online}
\end{figure*}
\begin{figure*}[bp] 
   \centering
   \epsscale{1.2}
   \plotone{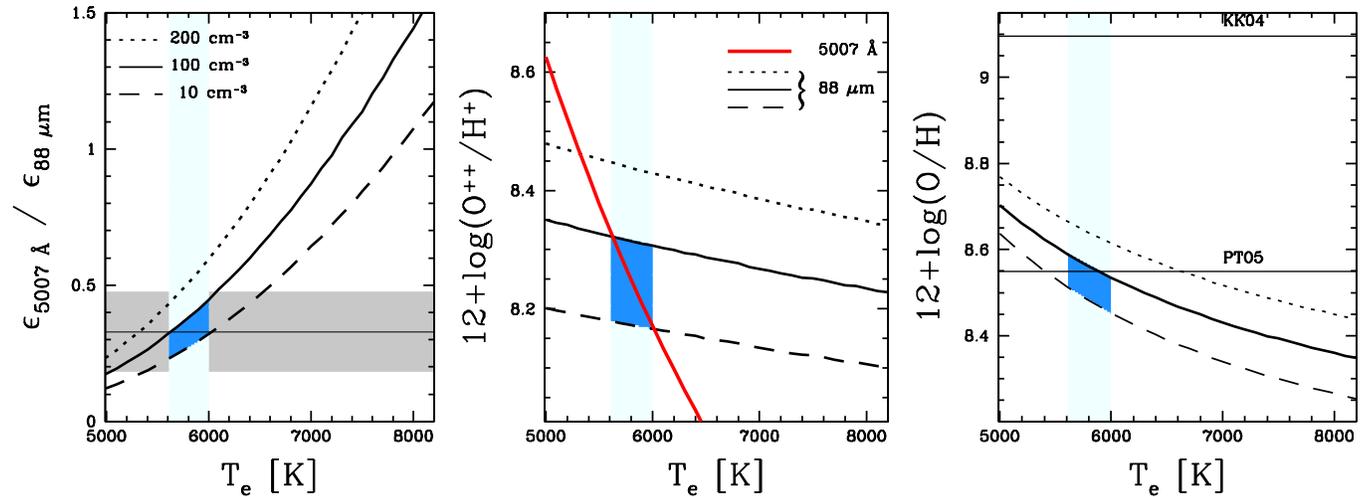}
   \caption{ Region 3.}
   \label{fig:region_online}
\end{figure*}
\begin{figure*}[bp] 
   \centering
   \epsscale{1.2}
   \plotone{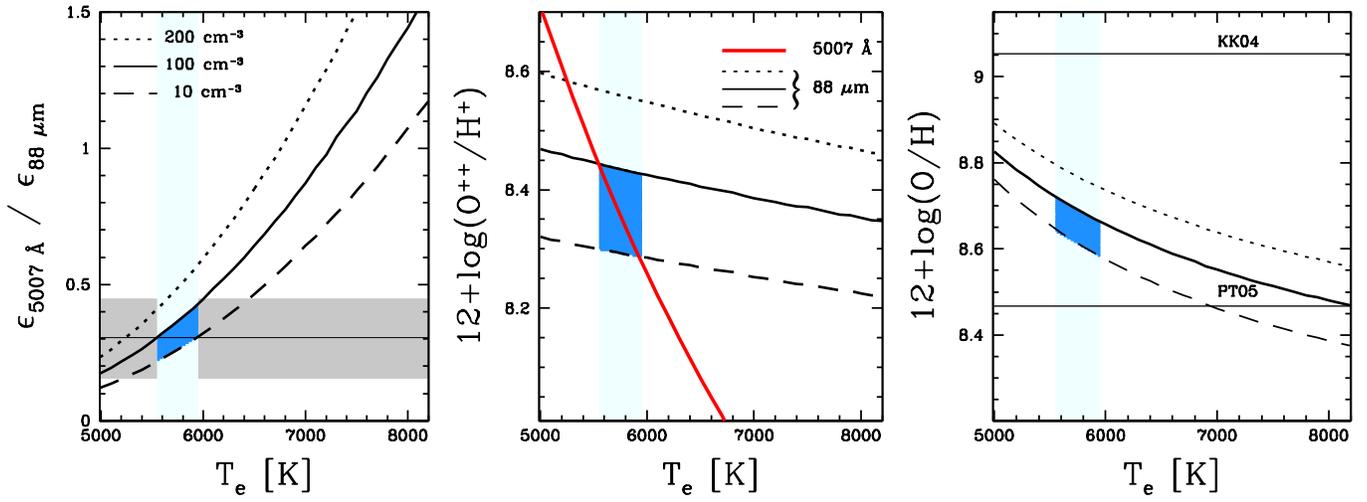}
   \caption{ Region 4.}
   \label{fig:region_online}
\end{figure*}
\begin{figure*}[bp] 
   \centering
   \epsscale{1.2}
   \plotone{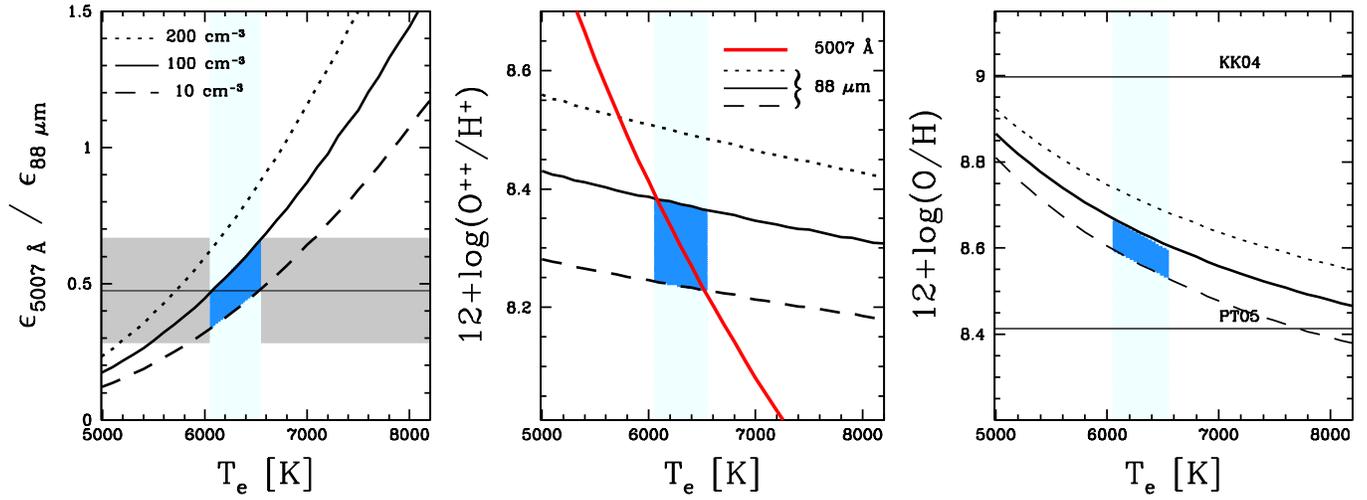}
   \caption{ Region 5.}
   \label{fig:region_online}
\end{figure*}
\begin{figure*}[bp] 
   \centering
   \epsscale{1.2}
   \plotone{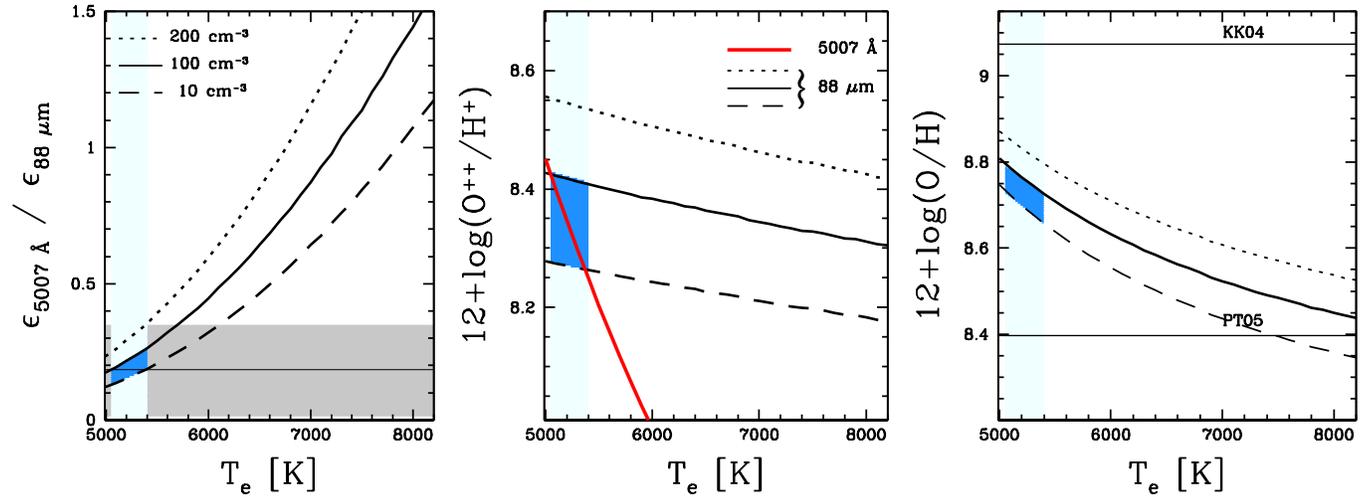}
   \caption{ Region 7.}
   \label{fig:region_online}
\end{figure*}

\end{document}